# A Novel Fractional Order Fuzzy PID Controller and Its Optimal Time Domain Tuning Based on Integral Performance Indices

Saptarshi Das[a,b], Indranil Pan[b], Shantanu Das[c] and Amitava Gupta[a,b]

a) School of Nuclear Studies & Applications (SNSA), Jadavpur University, Salt Lake Campus, LB-8, Sector 3, Kolkata-700098, India. Email: saptarshi@pe.jusl.ac.in, amitg@pe.jusl.ac.in

b) Department of Power Engineering, Jadavpur University, Salt Lake Campus, LB-8, Sector 3, Kolkata-700098, India. Email: indranil.jj@student.iitd.ac.in, indranil@pe.jusl.ac.in

c) Reactor Control Division, Bhabha Atomic Research Centre, Mumbai-4000854, India. Email: shantanu@magnum.barc.gov.in

**Abstract:**
A novel fractional order (FO) fuzzy Proportional-Integral-Derivative (PID) controller has been proposed in this paper which works on the closed loop error and its fractional derivative as the input and has a fractional integrator in its output. The fractional order differ-integrations in the proposed fuzzy logic controller (FLC) are kept as design variables along with the input-output scaling factors (SF) and are optimized with Genetic Algorithm (GA) while minimizing several integral error indices along with the control signal as the objective function. Simulations studies are carried out to control a delayed nonlinear process and an open loop unstable process with time delay. The closed loop performances and controller efforts in each case are compared with conventional PID, fuzzy PID and $PI^\lambda D^\mu$ controller subjected to different integral performance indices. Simulation results show that the proposed fractional order fuzzy PID controller outperforms the others in most cases.

**Index Terms:** Fractional order controller; fuzzy PID; FLC tuning; integral performance indices; Genetic Algorithm; optimal PID tuning.

## 1. Introduction:

So far the focus of the engineering community had been primarily on expressing systems with integer order differential equations and using a multitude of analytical and numerical solutions to optimize the formulation and analysis procedure. However recent developments in hardware implementation [1]-[3] of fractional order elements have brought a renewed wave in the modeling and analysis of new class of fractional order systems which look at natural phenomenon from a whole new perspective. The theory for fractional order systems have existed for the last 300 years [1]. These extend our common notion of integer order (IO) differential equations to include fractional powers in the derivative and integral terms and have been shown to model natural processes more accurately than IO differential equations. However the mathematical analysis behind these kinds of FO systems is naturally more involved than IO systems.

From classical control engineering perspective the stress has always been to obtain linearized model of a process and the controller as the control theory for these types of systems are already well formulated. With the advent of fuzzy set-theory there is perhaps some more flexibility in designing systems and expressing the observations in a more easy to follow linguistic notation. The fuzzy logic controller in a closed loop control system is basically a static non-linearity between its inputs and outputs, which can be tuned easily to match the desired performance of the control system in a more heuristic manner without delving into the exact mathematical description of the modeled nonlinearity. Traditional PID controllers work on the basis of the inputs of error, the derivative and the integral of error. An attempt can be made to justify the logic of incorporating a fractional rate of error as an input to a controller instead of a pure derivative term. Assuming that a human operator replaces the automatic controller in the closed loop feedback system, the human operator would rely on his intuition, experience and practice to formulate a control strategy and he would not do the differentiation and integration in a mathematical sense. However the controller output generated as a result of his actions may be approximated by appropriate mathematical operations which have the required compensation characteristics. Herein lies the applicability of FO derivatives or integrals over their IO counterparts as better approximation of such type of control signals, since it gives additional flexibility to the design. The rationale behind incorporating fractional order operators in the FLC input and output can be visualized like an heuristic reasoning for an observation of a particular rate of change in error (not in mathematical sense) by a human operator and the corresponding actions he takes over time which is not static in nature since the fractional differ-integration involves the past history of the integrand and as if the integrand is continuously changing over time [1]. Since the human brain does not observe the rate of change of a variable and its time evolution as pure numerical differentiation and integration, the fractional orders of differ-integration perhaps put some extra flexibility to map information in a more easily decipherable form. Our present day's mathematical modeling techniques, motivated by integer order differ-integrals does not give this flexibility and fails to describe it adequately. It is investigated in the present study that the fractional rate of error perhaps is capable of providing extra flexibility in the design of conventional FLC based PID controllers [4] which works with error and its rate in a pure mathematical sense (IO). It is logical that the fractional rate of error introduces some extra degree of flexibility in the input variables of FLC and can be tuned also like the input-output scaling factors as the FLC gain and shape of the membership functions (MF) to get enhanced closed loop performance. The present study investigates the effectiveness of the proposed fuzzy FOPID controller at producing better performance compared to classical PID, fuzzy PID and even $PI^\lambda D^\mu$ controllers due to higher degrees of freedom for tuning. However, the objectives to be fulfilled by different controller structures must be chosen judiciously.

In this paper, the parameters of this new kind of fractional order fuzzy logic controller are optimally tuned with GA to handle a delayed nonlinear process and an open loop unstable process with time delay. Time domain performances of other controller structures viz. PID, fuzzy PID and $PI^\lambda D^\mu$ using Integral of Squared Control Signal (ISCO) along with various integral error indices like Integral of Time multiplied Absolute Error (ITAE), Integral of Time multiplied Squared Error (ITSE), Integral of Squared Time multiplied Error whole Squared (ISTES) and Integral of Squared Time multiplied

Squared Error (ISTSE) are compared and the effectiveness of the different controllers are evaluated therein. Optimal tuning of FLC based PID can be found in few literatures. Hu, Mann & Gosine [5] tuned the FLC MFs along with the input-output SFs using GA to minimize a weighted summation of Integral of Squared Error (ISE) normalized by maximum error, maximum percentage of overshoot and settling time normalized by simulation time. Woo, Chung & Lin [6] have shown that tuning of MFs have lesser effect on the closed loop performance than the input-output SFs of a fuzzy PID controller. Their relative impact can be viewed like changing the universe of discourse for fuzzy inference by the input SFs and amplifying the defuzzified control law by output SFs while acting as the conventional PID controller gains. Also, Pan, Das & Gupta [7] designed an optimal fuzzy PID controller by minimizing the ITAE and ISCO to handle the effect of random delays in networked control systems (NCS). The present study assumes fixed MFs and rule base for the FLC as in its IO counterpart [6] and then tunes the fractional rate of error, fractional order integration of FLC output along with the input-output SFs to achieve optimum performance in time domain i.e. low control signal and error index.

The rest of the paper is organized as follows. Section 2 gives a brief review of the existing intelligent techniques for designing fractional order controllers and introduces the novelty of the proposition in the present study. Section 3 describes the structure of the fractional order fuzzy PID controller with the details of the rule base and the membership functions. The objective functions (time domain performance indices) along with genetic algorithm that has been used for the optimization are introduced in this section. Section 4 gives a comparison of the simulation results for two different class of processes. The paper ends with the conclusions in Section 5, followed by the references.

## 2. Review of the existing intelligent tuning techniques of FO controllers:

Classical notion of PID controllers has been extended to a more flexible structure $PI^\lambda D^\mu$ by Podlubny [8] with the fractional differ-integrals as the design variables along with the controller gains. Several intelligent techniques have been proposed for efficient tuning of such fractional order $PI^\lambda D^\mu$ controllers. Dominant pole placement based optimization problems have been attempted to design $PI^\lambda D^\mu$ controllers using Differential Evolution in Maiti *et al.* [9], Biswas *et al.* [10] and Invasive Weed Optimization with Stochastic Selection (IWOSS) in Kundu *et al.* [11]. Maiti *et al.* [12] also tuned a FOPID controller for stable minimum phase systems by minimizing ITAE criteria with Particle Swarm Optimization (PSO). A similar approach has been adopted for optimization of a weighted sum of Integral of Absolute Error (IAE) and ISCO to find out the controller parameters with GA by Cao, Liang & Cao [13] and with PSO by Cao & Cao [14]. Cai, Pan & Du [15] tuned a $PI^\lambda D^\mu$ controller by minimizing the ITAE criteria using multi-parent crossover evolutionary algorithm. Luo & Li [16] tuned a similar ITAE based $PI^\lambda D^\mu$ controller with Bacterial Foraging oriented by Particle Swarm Optimization (BF-PSO). Meng & Xue [17] designed a $PI^\lambda D^\mu$ controller using a multi-objective GA which minimizes the infinity-norm of the sensitivity (load disturbance suppression), and complementary sensitivity function (high frequency measurement noise rejection), rise time ($t_r$) and percentage of maximum overshoot ($M_p$) and additionally meets the specified gain cross-over frequency ($\omega_{gc}$), phase margin ($\phi_m$) and iso-damping property rather than minimizing these as a single objective with a weighted summation like Zamani *et al.* [18]. Dorcak *et al.* [19] proposed a frequency domain robust $PI^\lambda D^\mu$ controller tuning methodology using Self-Organizing Migrating Algorithm (SOMA), which is an extension of that proposed by Monje *et al.* [20] using constrained Nelder-Mead Simplex algorithm. Zhao *et al.* [21] tuned a $PI^\lambda D^\mu$ controller for inter-area oscillations in power systems by minimizing a weighted sum of the $M_p$, settling time ($t_s$) and error signal ($e$) using a GAPSO algorithm. Kadiyala, Jatoth & Pothalaiah [22] designed PSO based optimization problem for minimizing a weighted sum of $t_r, M_p, t_s$, steady-state error ($e_{ss}$) to design a $PI^\lambda D^\mu$ controller for aerofin control system. A PSO based similar approach can be found in Sadati, Zamani & Mohajerin [23] for SISO and MIMO systems. Sadati, Ghaffarkhah & Ostadabbas [24] designed a Neural Network based FOPID controller by minimizing the Mean Square Error (MSE) of the closed loop system while weights of the Neural Network and fractional orders are determined in the learning phase and the controller gains are adapted with change in the error. Ou, Song & Chang [25] designed a FOPID controller for First Order Plus Time Delay (FOPTD) systems using Radial Basis Function (RBF) neural network where the controller gains and differ-integrals can be determined from the time constant and delay of the process after the neural network is trained with a large set of FOPID parameters and system parameters with available frequency domain robust tuning methods. Weighted sum of several time-domain and frequency-domain criteria based optimization approach has been used to tune a FOPID controller with PSO for an automatic voltage regulator by Ghartemani *et al.* [26] and Zamani *et al.* [18], [27]. The approach in [27] also proposes an $H_\infty$-optimal FOPID controller by putting the infinity norm of the weighted sensitivity and complementary sensitivity functions as an inequality constraint to the objective function that in [18]. Lee & Chang [28], [29] used Improved ElectroMagnetism with Genetic Algorithm (IEMGA) to minimize the Integral of Squared Error (ISE) while searching for optimal $PI^\lambda D^\mu$ parameters. Pan *et al.* [30] used evolutionary algorithms for time domain tuning of $PI^\lambda D^\mu$ controllers to cope with the network induced packet drops and stochastic delays in NCS applications.

Recent advent of few non-PID type intelligent fractional order controllers have been shown to be more effective over the existing technologies. Efe [31] used fractional order integration while designing an Adaptive Neuro-Fuzzy Inference System (ANFIS) based sliding mode control. Delavari *et al.* [32] proposed a fuzzy fractional sliding mode controller and tuned its parameters with GA. Barbosa *et al.* [33] incorporated fuzzy reasoning in fractional order PD controllers. Arena *et al.* [34]-[35] introduced a new Cellular Neural Network (CNN) with FO cells and studied existence of chaos in it. Valerio & Sa da Costa [36] studied fuzzy logic based approximation of variable complex and real order derivatives with and without memory.

In the present study, the tuning of a new fuzzy FOPID controller has been attempted with GA and the closed loop performances are compared with an optimal $PI^\lambda D^\mu$ controller. The input-output SFs and differ-integrals of the FO fuzzy PID controller are tuned while minimizing weighted sum of various error indices and control signal similar to that in Cao, Liang & Cao [13] and Cao & Cao [14] with a simple ISE criteria. While [31], [32] focuses on fractional order fuzzy sliding mode controllers, the present work is concerned with the fuzzy analogue of the conventional PID controller which is widely used in the process control industry. In Barbosa *et al.* [33], the fractional fuzzy PD controller is investigated in terms of digital implementation and robustness. However the tuning methodology is complex and might not always ensure optimal time domain performance. The performance improvement is even more for complicated and ill behaved systems which have been enforced to obey a set of desired control objective with GA in the present formulation.

## 3. New fractional order fuzzy PID controller and its time domain optimal tuning:
### 3.1. Structure of fractional order fuzzy PID controller:

The structure of the fuzzy PID used here is inherited from a combination of fuzzy PI and fuzzy PD controllers [4] with $K_e$ and $K_d$ as the input SFs and $\alpha$ and $\beta$ as output SFs as described by Woo, Chung & Lin [6] and Yesil, Guzelkaya & Eksin [37]. Typical advantage of this particular controller structure over other available FLC based PIDs are described in a detailed manner by Pan *et al.* [7]. However, in the original IO fuzzy PID controller the inputs were the error and the derivative of error and the FLC output was multiplied by $\alpha$ and its integral multiplied with $\beta$ and then summed to give the total controller output. But in the present case the integer order rate of the error at the input to the FLC is replaced by its fractional order counterpart ($\mu$). Also the order of the integral is replaced by a fractional order ($\lambda$) at the output of the FLC representing a FO summation (integration) of the FLC outputs. The values of these orders $\{\lambda, \mu\}$ along with $\{K_e, K_d, \alpha, \beta\}$ are the optimization variables in genetic algorithm.

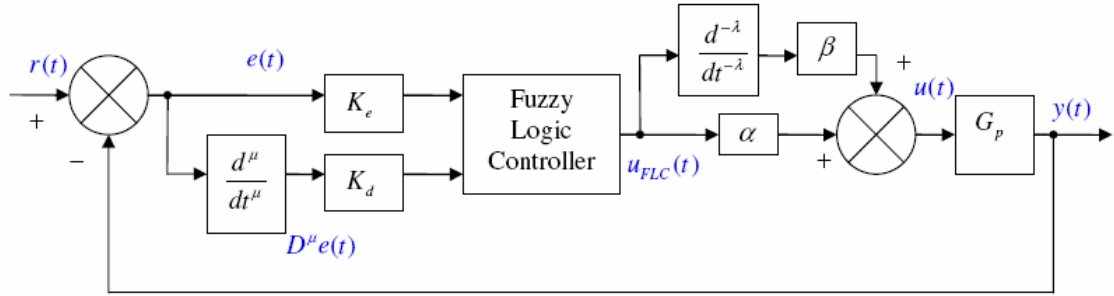

Fig. 1. Structure of the fractional order fuzzy PID controller.

For performance comparison a FOPID controller is also considered which has the transfer function in non-interacting or parallel structure:

$$C^{FOPID}(s) = K_p + \frac{K_i}{s^\lambda} + K_d s^\mu \tag{1}$$

The other controller structures like conventional integer order FLC and PID controller, used in the simulations can be obtained by simply setting the values of $\{\lambda, \mu\} = 1$ in Fig. 1 and equation (1) respectively. The proposed FLC based FOPID controller uses a two dimensional linear rule base (Fig 2) for the error, fractional rate of error and the FLC output with standard triangular membership functions and Mamdani type inferencing. In the present work the fuzzy rule base is derived using intuitive logic as in [4]. Mudi & Pal [4] has given a detailed reference as to how the fuzzy linguistic rules are translated into the inferencing process and is well accepted in the scientific community. The triangular membership function is chosen over the other types like Gaussian, Trapezoidal, Bell-shaped, π-shaped etc. as it is easier to implement in practical hardware [6], [7], [37]. In Fig. 2-3, the fuzzy linguistic variables NL, NM, NS, ZR, PS, PM, PL represent Negative Large, Negative Medium, Negative Small, Zero, Positive Small, Positive Medium and Positive Large respectively. The FLC output ($u_{FLC}$) is determined by using center of gravity method by defuzzification.

The membership functions (Fig. 3) and rule bases (Fig. 2) are similar to that in Woo, Chung & Lin [6] and Yesil, Guzelkaya & Eksin [37] for integer order FLC based PID controllers. For fuzzy controllers it is well known [4] that change in output scaling factor, for example, has more effect on the controller performance than changes in the shape of the membership functions. Thus all the tuning parameters are not equally potent in affecting the overall performance of the controller. The focus of the present paper is to investigate the effect of tuning the fractional rate of error ($\mu$), while keeping the rule base and shape of membership functions unchanged to enhance the overall closed loop performance of a control system. From the point of applicability and ease of use for the practicing engineer, this approach is better since the performance of the control system can be drastically affected by tuning these two additional parameters appropriately instead of the membership functions and other fuzzy inferencing variables. Fig. 4 shows the nonlinear surface plot for the rule base of the fuzzy logic controller.

The set of rules can be divided into the following five groups to understand the logic of incorporating the rule base as in Fig. 2, as detailed in [38].

*Group 0:*

In this group of rules the error ($e$) and its fractional derivative ($\frac{d^\mu e}{dt^\mu}$) have very small positive or negative values or are equal to zero. This implies that the process output has strayed off slightly from the set point but is still close to it. Hence small values of control signals are required to correct these small deviations and these rules mainly relate to the steady state behaviour of the process.

*Group 1:*

In this group the error is negative large or medium, implying that the process output is significantly above the setpoint. Also the fractional derivative of the error is positive implying that the process output is moving towards the set point. Thus the controller applies a signal to speed up or slow down the approach towards the set point.

*Group 2:*

In this group the error is either close to the set point (NS, ZR, PS) or is significantly below it (PM, PL). Also since the fractional rate of error is positive, the process output is moving away from the set point. Hence different levels of positive control signal is required for different combinations of $e$ and $\frac{d^\mu e}{dt^\mu}$, to reverse the course of the process output and make it tend towards the set point.

*Group 3:*

For this group $e$ is positive medium or big, suggesting that the process output is far below the set point. Simultaneously, since $\dfrac{d^{\mu}e}{dt^{\mu}}$ is negative, the process output is moving towards the set point. Hence the controller applies an appropriate signal, to speed up or slow down the approach towards the set point.

*Group 4:*
In this group the error is either close to the set point (NS, ZR, PS) or is significantly above it (NM, NL). Also since $\dfrac{d^{\mu}e}{dt^{\mu}}$ is negative, the process output variable is moving away from the set point. Hence a negative control signal reverses this trend and tries to make the process output change move towards the set point.

Also in this case only 7 linguistic variables have been used resulting in 49 rules in the table. Since the linguistic variables dictate the granularity of the control action, more number of them could be used for better control resolution. But in these cases the rule base increases in the order of $n^2$ (where $n$ is the number of linguistic variables) and hence would be difficult to implement in real time hardware.

Looking from the perspectives of computational efficiency, good memory usage and performance analysis requirements, a uniform representation of the membership function is sought after. Generally triangular, trapezoidal and bell shaped functions are preferred since their functional description can be easily obtained, they can be stored with minimal usage of memory and can be manipulated efficiently by the inference engine to meet the hard limits of real time requirements. Since the parametric, functional description of the triangular function is the most economic among these [38], it is widely adopted in controller design for real time applications and has been chosen in the present study.

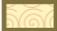

Fig. 2. Rule base for error, fractional rate of error and FLC output.

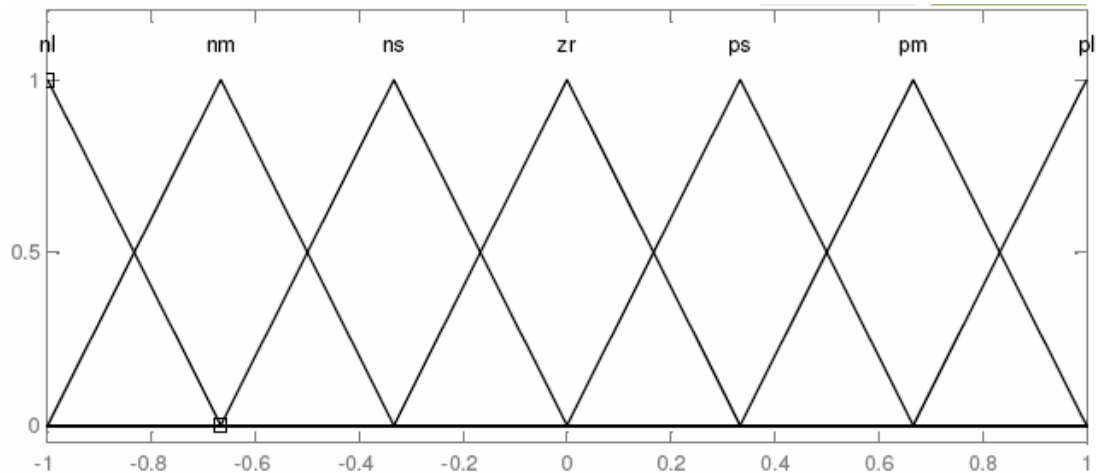

Fig. 3. Membership functions for error, fractional rate of error and FLC output.

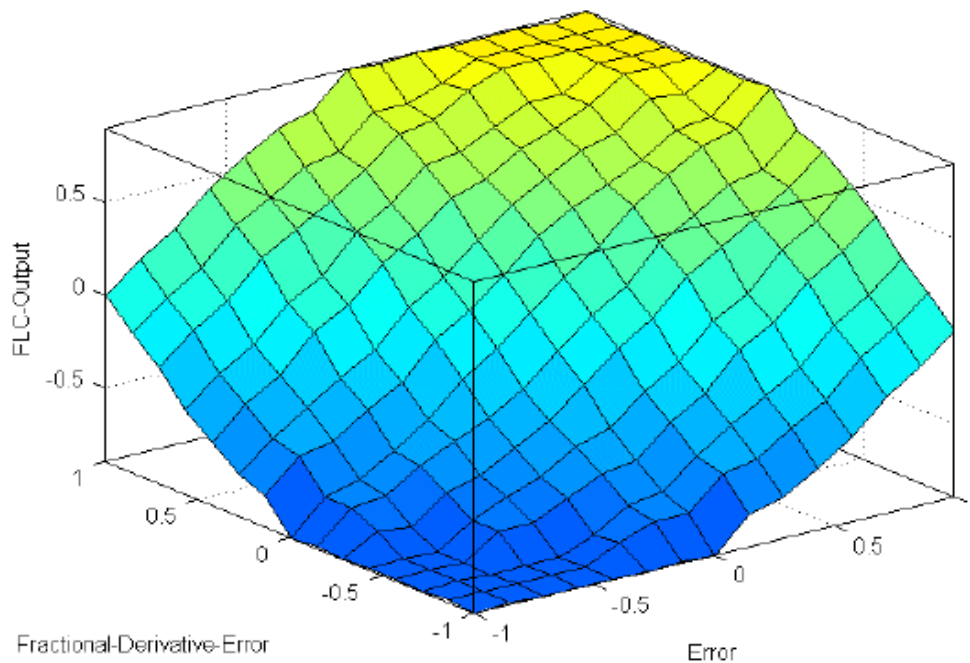

Fig. 4. Surface plot for rule base.

Also, there are several continuous and discrete time rational approximation methods available for fractional order elements [1]-[3], [39]-[41]. In the present simulation study, each guess value of the fractional order differ-integrals $\{\lambda, \mu\}$ within the optimization process are continuously rationalized with Oustaloup's 5$^{th}$ order rational approximation [42].

The fractional order differ-integrals are basically infinite dimensional linear filters. However band-limited realization of $PI^\lambda D^\mu$ controllers is necessary for its hardware implementation. In the present simulation study each fractional order element has been rationalized with Oustaloup's recursive filter [42] given by the following equation (2)-(3). If it be assumed that the expected fitting range or frequency range of controller operation is $\left(\omega_b, \omega_h\right)$, then the higher order filter which approximates the FO element $s^\gamma$ can be written as:

$$G_f(s) = s^\gamma = K \prod_{k=-N}^{N} \frac{s + \omega_k'}{s + \omega_k} \tag{2}$$

where the poles, zeros, and gain of the filter can be evaluated as:

$$\omega_k = \omega_b \left(\frac{\omega_h}{\omega_b}\right)^{\frac{k+N+\frac{1}{2}(1+\gamma)}{2N+1}}, \omega_k' = \omega_b \left(\frac{\omega_h}{\omega_b}\right)^{\frac{k+N+\frac{1}{2}(1-\gamma)}{2N+1}}, K = \omega_h^\gamma \tag{3}$$

In (2) and (3), $\gamma$ is the order of the differ-integration and $\left(2N+1\right)$ is the order of the filter. Present study considers a 5th order Oustaloup's rational approximation [42] for the FO elements within the frequency range $\omega \in \left\{10^{-2}, 10^2\right\}$ rad/sec which is most common in process control applications.

### 3.2. Formulation of the objective function for time domain optimal controller tuning:

Various time domain integral performance indices like ITAE, ITSE, ISTES and ISTSE are considered in the problem similar to that in [43]. Tavazoei [44] studied finiteness of such integral performance indices for fractional order systems with unit step set-point and load-disturbance excitation. Zhuang & Atherton [45] first proposed optimization of such integral performance indices for time domain tuning of PID controllers which is extended for $PI^\lambda D^\mu$ controllers by Cao, Liang & Cao [13] and Cao & Cao [14] and further for FO fuzzy PID controllers in the present paper.

Every integral performance index has certain advantages in control system design. The ITAE criterion tries to minimize time multiplied absolute error of the control system. The time multiplication term penalizes the error more at the later stages than at the beginning and hence effectively reduces the settling time ($t_s$) which can not be achieved with IAE or ISE based tuning [5, 13, 14, 28, 29]. Since the absolute error is included in the ITAE criterion, the maximum percentage of overshoot ($M_p$) is also minimized. The ITSE criterion penalizes the error more than the ITAE and due to the time multiplication term, the oscillation damps out faster. However for a sudden change in set-point the ITSE based controller produces larger controller output than the ITAE based controllers, which is not desirable from actuator design point of view. Other integral performance indices like ISTES and ISTSE both have higher powers of time and error terms. These result in faster rise time and settling time while also ensuring the minimization of the peak overshoot. These however might lead to very high value of control signal and are only used in acute cases where the time domain performance is of critical importance and not a large control signal.

To avoid large control signal which may saturate the actuator and cause integral wind-up, it is also minimized as a part of the objective function with GA. The objective functions used for controller tuning has been taken as the weighted sum of several performance indices along with the controller outputs as shown below. The weights $\left\{w_1, w_2\right\}$ in the control objective (4)-(7) gives extra flexibility to the designer, depending on the nature of application and relative importance of low error index and low control signal. The four objectives, used in the present simulation study are presented in (2)-(5).

$$J_1 = \int_0^\infty \left[ w_1 \cdot t \left| e(t) \right| + w_2 \cdot u^2(t) \right] dt = \left( w_1 \times ITAE \right) + \left( w_2 \times ISCO \right) \tag{4}$$

$$J_2 = \int_0^\infty \left[ w_1 \cdot te^2(t) + w_2 \cdot u^2(t) \right] dt = \left( w_1 \times ITSE \right) + \left( w_2 \times ISCO \right) \tag{5}$$

$$J_3 = \int_0^\infty \left[ w_1 \cdot \left( t^2 e(t) \right)^2 + w_2 \cdot u^2(t) \right] dt = \left( w_1 \times ISTES \right) + \left( w_2 \times ISCO \right) \tag{6}$$

$$J_4 = \int_0^\infty \left[ w_1 \cdot t^2 e^2(t) + w_2 \cdot u^2(t) \right] dt = \left( w_1 \times ISTSE \right) + \left( w_2 \times ISCO \right) \tag{7}$$

Optimization is carried out to obtain controller parameters, considering equal weights for integral error indices and integral of control signal i.e. $w_1 = w_2$. Now, minimization of the control objectives (4)-(7), gives optimal parameters for the FLC based FOPID and the $PI^\lambda D^\mu$ controller. Optimal parameters for much simpler controllers i.e. fuzzy PID and PID are also searched by putting the FO differ-integrals as unity. A large penalty function has been incorporated in the optimization process for very large value of $J$ to avoid parameter search with unstable closed loop response as suggested by Zamani *et al.* [18].

From control system designer's point of view, the ISE based PID controller designs are popular since using Parseval's theorem, this typical time domain performance index represents the $H_2$-norm of the closed loop system in frequency domain. It is well known that analytical stability study with the error model using ISE ($H_2$-norm) criteria is possible only for linear system. In the present framework, the only method to cope with the nonlinearities due to the process model itself as well as the fuzzy inference to ensure well behaved control performance is to optimize time domain error indices involving time along with the associated cost of control. However, the frequency domain analysis is difficult with higher powers of error and time (like ITSE, ISTES, ISTSE) for the integral performance index and even more with the addition of a linear/nonlinear control law with the error index. The proposed family of time domain integral performance indices based tuning technique is especially needed for processes, governed by highly nonlinear differential equations and not mere linear systems with actuator nonlinearities, commonly encountered in process controls.

### *3.3. Optimization algorithm used for the tuning of optimal controllers:*

Gradient based classical optimization algorithms for minimization of the objective function often get trapped in the local minimas. This can be overcome by any stochastic algorithms like Genetic Algorithm which has been used here to find the optimum set of values for the controller parameters. The variables that constitute the search space for the fractional fuzzy PID controller are $\left\{ K_e, K_d, \alpha, \beta, \lambda, \mu \right\}$. The intervals of the search space for these variables are $\left\{ K_e, K_d, \alpha, \beta \right\} \in \left[ 0, 100 \right]$ and $\left\{ \lambda, \mu \right\} \in \left[ 0, 2 \right]$. The variables are encoded as real values in the algorithm.

Genetic algorithm is a stochastic optimization process inspired by natural evolution. During the initialization phase, a random population of solution vectors with uniform distribution is created over the whole solution domain. The population is encoded as a double vector and the bit string representation is not used. A feasible population set which would always give stable controller outputs was not chosen since the linear or nonlinear constraints for stability is difficult to derive analytically in the present case with process and controller nonlinearities in the loop. This implies that some of the controller parameters may produce unstable closed loop response and terminate the code due to this ill-conditioning. Thus to overcome this problem, whenever the time evolution of the objective function shows instability, a large value of the objective function is assigned (10000 for this case) without simulating it for the entire time horizon. This automatically assigns a bad fitness rank to the solution and the unstable modes are eliminated over successive generations. Each solution vector in the present population undergoes reproduction, crossover and mutation stochastically, in each generation, to produce a better population of solution vectors (in terms of fitness values) in the next generation. A scaling function converts the raw fitness scores in a form that is suitable for the selection function. Various scaling functions like rank, proportional, top, shift linear scaling may be used. In this case rank fitness scaling is used which scales the raw scores on the basis of its position in the sorted score list. This removes the effect of the spread of the raw scores.

The process where solution vectors with higher fitness values can produce more copies of themselves in the next generation is known as reproduction. The number of fittest individuals (solution vectors) that will definitely be self replicated to the next generation is denoted in the algorithm by a parameter called the elite count. Increasing the elite count may result in domination of the fitter individuals obtained earlier in the simulation process. This will result in less effective solutions as the exploration of the search space would be limited. Thus, the parameter is generally a small fraction of the total population size. In this case, the population size is considered to be 20 and elite count as 2.

Crossover is the process in which two randomly selected vectors from the current generation of the gene pool, undergo an information exchange of probabilistic nature, to give rise to better individuals in the next generation. The crossover function may be scattered, single point, two point, heuristic, arithmetic etc. which are basically different mathematical operations in which the child can be created from the parent genes. In this case a scattered crossover function is used which creates a random binary vector and selects the genes where the vector has a value of 1 from the first parent, and the genes where the vector has a value of 0 from the second parent, and combines the genes to form the child. A user specified cross over fraction determines what percentage of the population (other than the elite) evolves through crossover. The remaining fraction, evolve through mutation. In mutation a small randomly chosen part of a solution vector is probabilistically altered to give rise to the child in the next generation. The mutation operation can be implemented with various mutation functions like Gaussian, uniform, adaptive feasible etc. which like the crossover function, are a set of mathematical operations that dictates how the mutated child will be formed from the parents. For mutation, in this case, the Gaussian function is used which adds a random number to each vector entry of an individual. This random number is taken from a Gaussian distribution centered around zero. The algorithm refines the solutions in this way iteratively until the change in the objective function is less than a pre-specified tolerance level or the maximum number of iterations is exceeded. If the crossover fraction is set to unity, there is no mutation and the algorithm stagnates after forming the best individual from the available gene pool. The best individual is successively replicated and carried forward through the generations without any improvement due to lack of mutation. Also if the crossover is set to zero and the population evolves purely through mutation, then this strategy improves the fitness of other individuals, but since these are never combined with the genes of the best individual due to lack of crossover, the best fitness levels off after sometime and the program terminates when the maximum number of iterations are exceeded. Hence, a judicious choice of the crossover and mutation fraction needs to be used. In the present optimization framework, a crossover fraction of 0.8 and mutation fraction of 0.2 has been used which gives satisfactory results for a wide variety of problems [46]. The other parameters of GA like population size, scaling function, selection function, elite count, mutation function, crossover function, which are used in the simulations, are also chosen in the lines of the previous argument [46]. The selection function chooses the vectors which act as parents of the next generation based on the inputs from the fitness scaling function. Here a stochastic uniform function is used.

Here, GA progressively minimizes the objective functions (4)-(7) over the iterations while finding optimal set of parameters for the FO fuzzy PID controller. The program terminates if the value of the objective function does not change appreciably over consecutive iterations (i.e. the change is less than the prespecified tolerance level) or the maximum number of iterations are exceeded. The maximum number of iterations is kept as 100 and the tolerance level is kept as $10^{-6}$.

## 4. Simulations and Results:
### 4.1. Nonlinear process with time delay:

The optimal tuning of the proposed FO fuzzy PID and other three controllers viz. $PI^{\lambda}D^{\mu}$, fuzzy PID and PID controllers are now carried out for a nonlinear process ($P_1$) as studied by Mudi & Pal [4]

$$\frac{d^2y}{dt^2} + \frac{dy}{dt} + 0.25y^2 = u(t-0.5) \tag{8}$$

The objective functions (4)-(7) are minimized for each of the fuzzy enhanced and the nominal controllers with the corresponding controller parameters reported in Table 1 and 2 respectively. As discussed earlier the adopted search range for the FO fuzzy PID controller parameters are restricted to $\{K_e, K_d, \alpha, \beta\} \in [0,100]$ and $\{\lambda, \mu\} \in [0,2]$ and for the FOPID controller the search range is $\{K_p, K_i, K_d\} \in [0,100]$ and $\{\lambda, \mu\} \in [0,2]$. Low gain of fuzzy FOPID controller is desired to keep the control signal small and the actuator size since control signals are directly proportional with output scaling factors. Restricting the input scaling factors to unity is to ensure that the fuzzy inference is always between the designed universe of discourse. Differ-integral orders greater than 1 leads to improper transfer function upon rational approximation and thus has been divided in two parts for simulation as suggested in Das *et al.* [47]. Restricting the order of integral to 2, is due to the fact that double integrating open loop systems are inherently unstable.

Now, with the four objective functions (4)-(7) and the four set of optimal controllers, the time response curves and the control signals are compared to show the relative potential of each of the controllers and the integral performance indices as well. Fig 5 shows the time response of plant $P_1$ with ITAE based tuning for unit change in set point and load disturbance. The capability of set point tracking for IO and FO fuzzy PID controllers are better than the corresponding non-fuzzy controllers. But fuzzy PID gives slightly better results than all the others in this case for load disturbance suppression. It is closely followed by the FO fuzzy PID controller. The PID controller and the FOPID controller have larger peak overshoot and a poor load disturbance characteristic. Fig. 6 shows the controller output for this plant based on the ITAE tuning. The PID and the FOPID controllers have a larger initial controller output, while the controller outputs for the fuzzy PID and the fractional fuzzy PID are better.

Fig 7. shows the output with the load disturbance of plant $P_1$ with ITSE based tuning. The overshoot is much lesser than the ITAE based tuning as ITSE puts larger penalty to the error signal in the optimization process. The fuzzy FOPID shows the best load disturbance rejection in this case and also best set-point tracking. Fig. 8 shows the control signal of the different controllers. The fuzzy PID controllers have a higher value of initial control signal. The controller output of the PID and FOPID controllers are smaller. It is logical that ITSE based tuning gives better time response Fig. 7) for the most flexible controller structures but at the cost of increased control signal (Fig. 8), since the penalties on large errors increases for ITSE criteria.

Table 1:
Optimal parameters for fuzzy FOPID and fuzzy PID controller for plant $P_1$

| Controller type | Performance index | $J^{\min}$ | Controller parameters | | | | | |
|---|---|---|---|---|---|---|---|---|
| | | | $K_e$ | $K_d$ | $\alpha$ | $\beta$ | $\lambda$ | $\mu$ |
| Fuzzy FOPID | ITAE and ISCO | 5.52735 | 0.478803 | 0.605029 | 1.780246 | 0.865874 | 0.999794 | 0.999598 |
| | ITSE and ISCO | 4.423768 | 0.307997 | 0.363393 | 1.731677 | 0.661103 | 0.957083 | 0.908509 |
| | ISTES and ISCO | 6.478104 | 0.628164 | 0.735571 | 1.600304 | 0.712215 | 0.999858 | 0.993217 |
| | ISTSE and ISCO | 4.70575 | 0.59135 | 0.676432 | 1.586057 | 0.621792 | 0.993939 | 1.0 |
| Fuzzy PID | ITAE and ISCO | 5.375536 | 0.674181 | 0.847209 | 1.346672 | 0.690657 | - | - |
| | ITSE and ISCO | 4.445912 | 0.663763 | 0.684081 | 1.301122 | 0.315402 | - | - |
| | ISTES and ISCO | 6.12801 | 0.632049 | 0.755715 | 1.715719 | 0.789966 | - | - |
| | ISTSE and ISCO | 4.687693 | 0.651062 | 0.710916 | 1.506431 | 0.537904 | - | - |

Table 2:
Optimal parameters for FOPID and PID controller for plant $P_1$

| Controller type | Performance index | $J^{\min}$ | Controller parameters | | | | |
|---|---|---|---|---|---|---|---|
| | | | $K_p$ | $K_i$ | $K_d$ | $\lambda$ | $\mu$ |
| FOPID | ITAE and ISCO | 6.936568 | 0.337983 | 0.155569 | 0.497122 | 0.972147 | 0.556586 |
| | ITSE and ISCO | 4.508684 | 0.085538 | 0.14587 | 0.56976 | 0.939418 | 0.346626 |
| | ISTES and ISCO | 13.35903 | 0.650325 | 0.191647 | 0.634971 | 0.989976 | 0.802389 |
| | ISTSE and ISCO | 5.007396 | 0.162653 | 0.176027 | 0.625217 | 0.946232 | 0.42833 |
| PID | ITAE and ISCO | 5.243994 | 0.962818 | 0.136967 | 0.924735 | - | - |
| | ITSE and ISCO | 4.426278 | 0.898051 | 0.114825 | 0.866315 | - | - |
| | ISTES and ISCO | 5.600617 | 1.285486 | 0.163374 | 1.083274 | - | - |
| | ISTSE and ISCO | 4.651827 | 1.080559 | 0.142924 | 1.014246 | - | - |

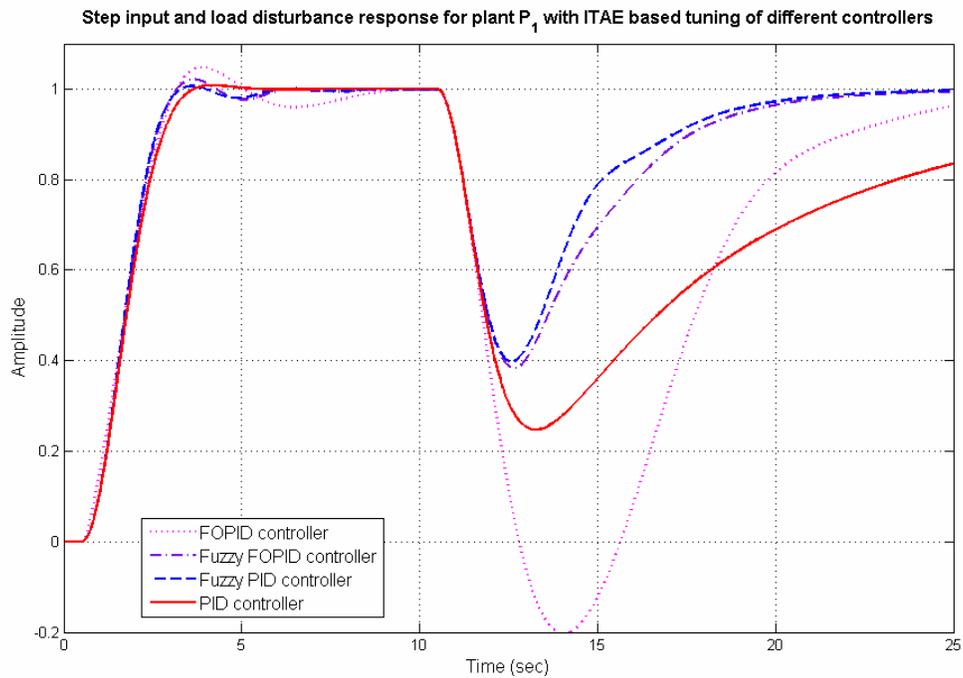

Fig. 5. Step input and load disturbance response for plant $P_1$ with ITAE based tuning.

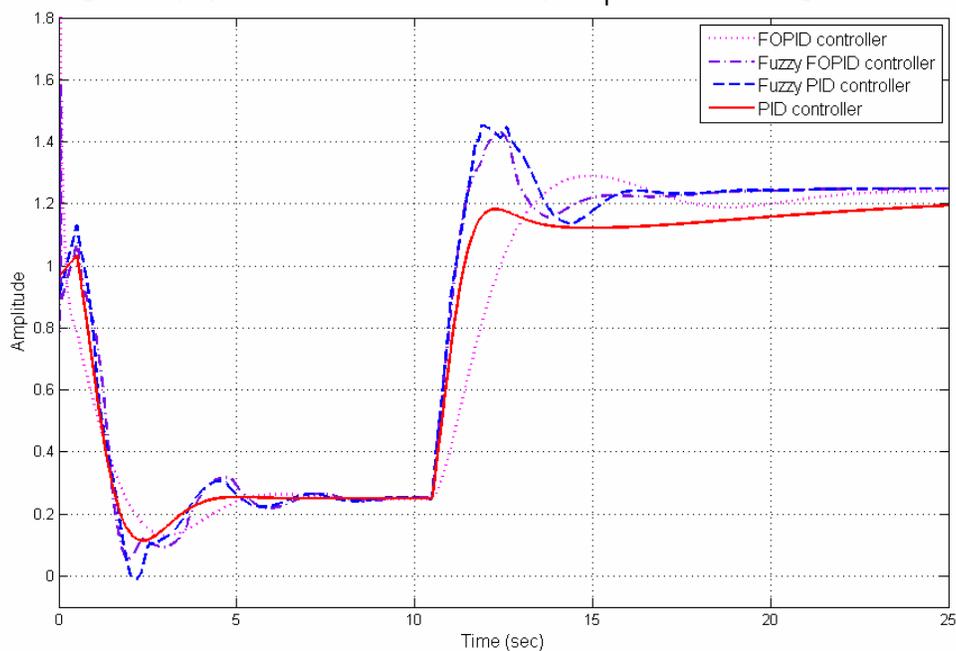

Fig. 6. Control signal of $P_1$ for step input and load disturbance with ITAE based tuning.

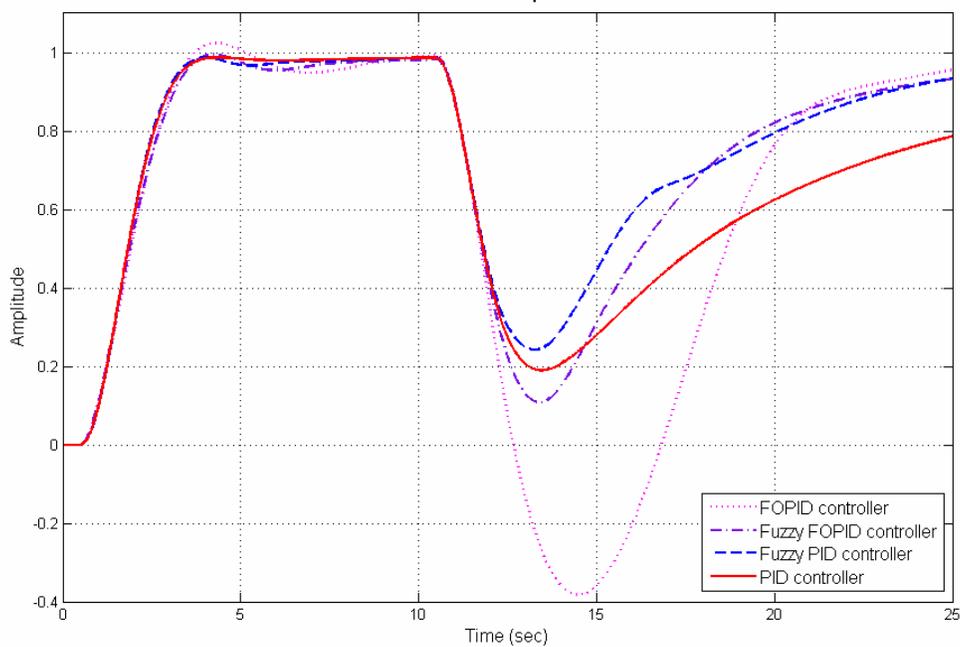

Fig. 7. Step input and load disturbance response for plant $P_1$ with ITSE based tuning.

Control signal for step input and load disturbance excitation for plant P₁ with ITSE based tuning of different controllers

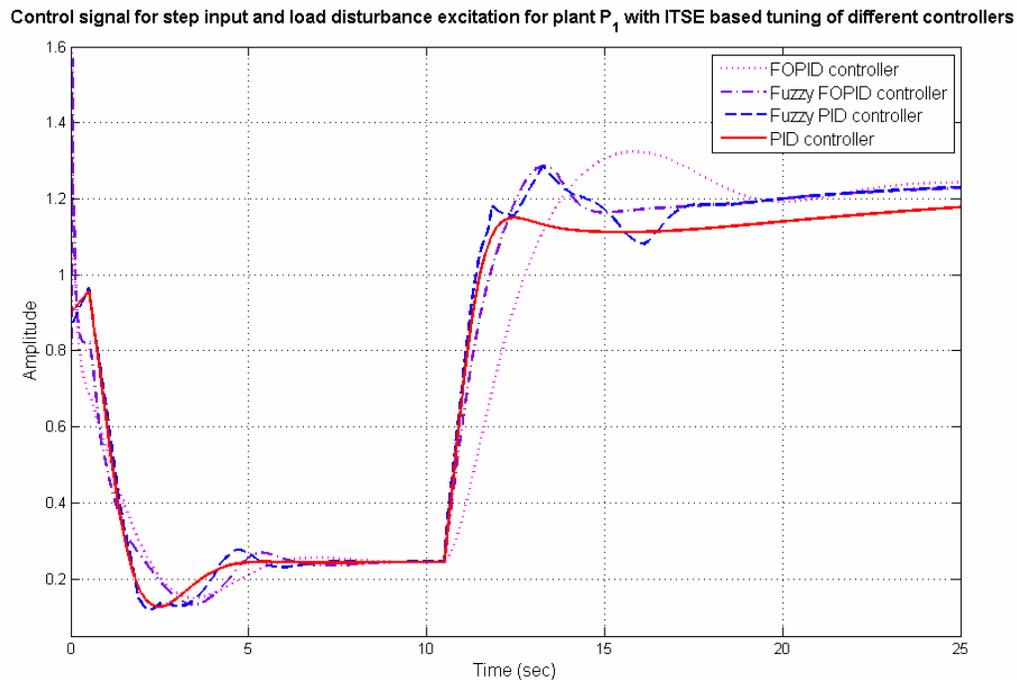

Fig. 8. Control signal of $P_1$ for step input and load disturbance with ITSE based tuning.

Step input and load disturbance response for plant P₁ with ISTES based tuning of different controllers

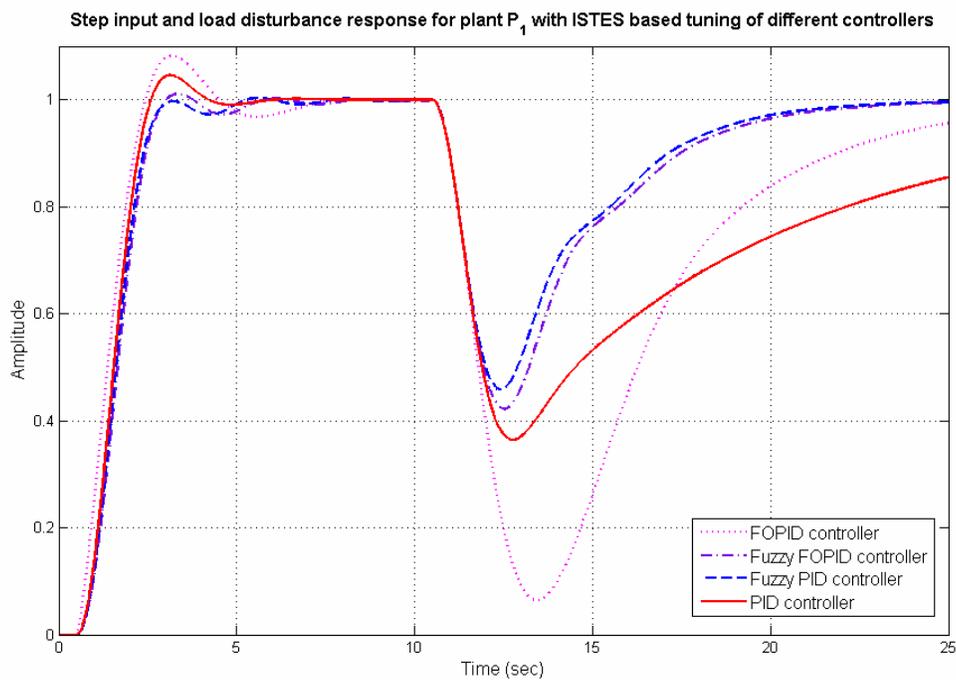

Fig. 9. Step input and load disturbance response for plant $P_1$ with ISTES based tuning.

Fig. 9 shows the unit step response of plant $P_1$ along with the load disturbance for ISTES criterion. The fuzzy PID and fuzzy FOPID both give a lower peak overshoot and a better load disturbance response. Fig. 10 shows the control signal for the ISTES criterion. The FOPID controller has a very large initial value which might result in actuator saturation. The fuzzy PID and fuzzy FOPID controllers have a relatively lower controller output.

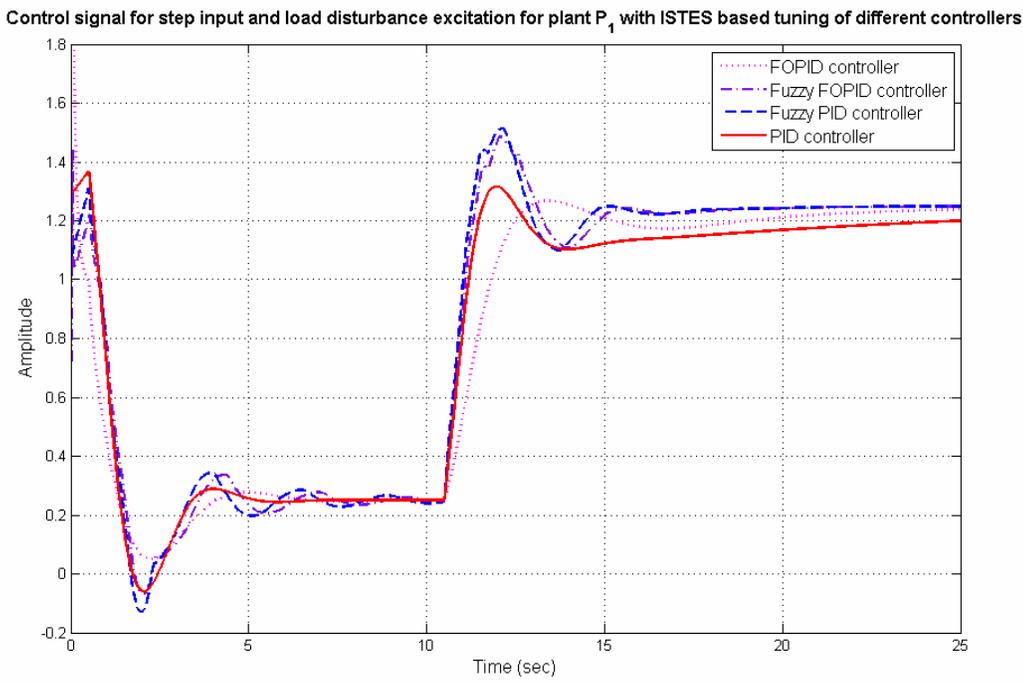

Fig. 10. Control signal of $P_1$ for step input and load disturbance with ISTES based tuning.

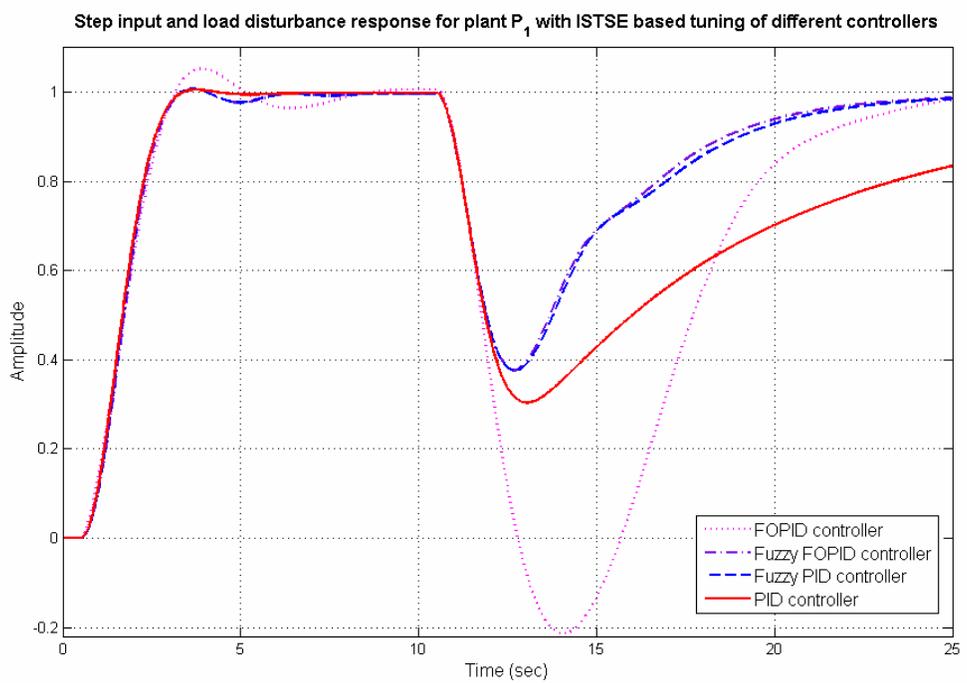

Fig. 11. Step input and load disturbance response for plant $P_1$ with ISTSE based tuning.

Fig. 11 shows the unit step response of plant $P_1$ along with the load disturbance for ISTSE criterion. The rise time is very sharp for all the controllers due to higher penalty on both time and error in the minimization criterion. The fuzzy PID gives the best load disturbance response closely followed by the fractional fuzzy PID controller. Fig. 12 shows the control signal output for this case. Both the fuzzy PID and FO fuzzy PID have lower value of initial control signal whereas it much higher for FOPID and PID.

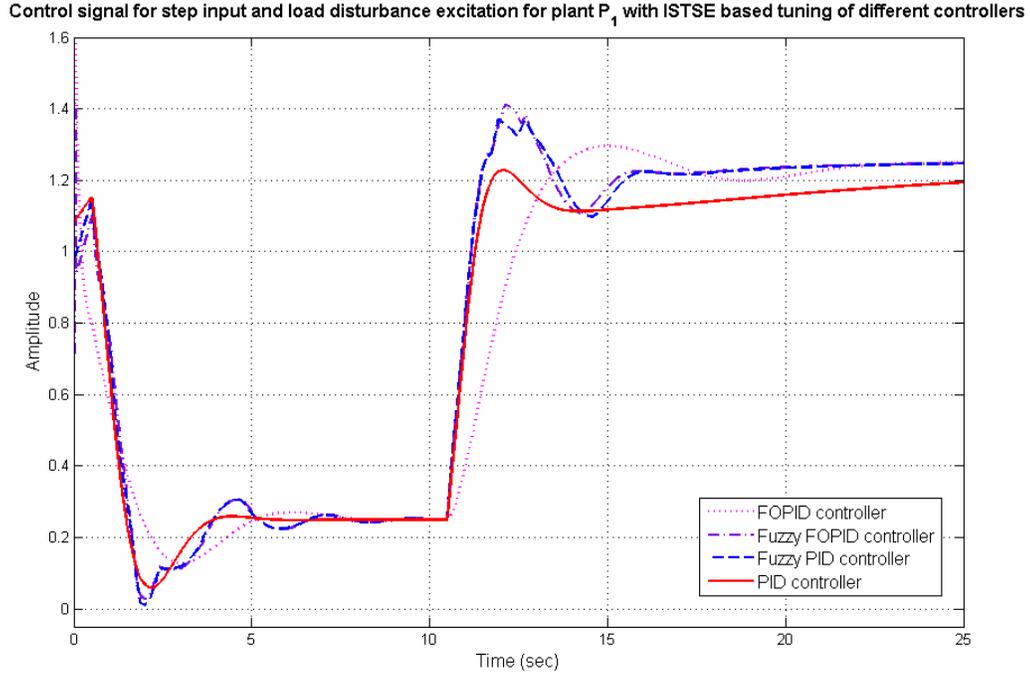

Fig. 12. Control signal of $P_1$ for step input and load disturbance with ISTSE based tuning.

### 4.2. Unstable process with time delay:

The next plant considered for performance study of the optimal controllers is that of an open loop unstable process with time delay as studied by Visioli [48].

$$P_2(s) = \frac{e^{-0.2s}}{(s-1)} \qquad (9)$$

Table 3 and 4 gives the optimal values of the controller parameters for the four different controllers, with four set of integral performance indices. Fig. 13 shows the unit step response of the plant $P_2$ along with the load disturbance for ITAE criterion. Both the fuzzy FOPID and the fuzzy PID controllers have almost no overshoot but the rise time of the fuzzy FOPID controller is better. The load disturbance suppression of the fuzzy FOPID is also better than the simple fuzzy PID controller. As is evident, the PID and FOPID controllers give high overshoot for delayed unstable processes. The time responses such a delayed unstable plant are much better with FLC based FOPID controller than that in the original work [48]. Fig. 14 shows the controller output for plant $P_2$ optimized with the ITAE criterion. The initial controller output of the PID and FOPID controllers is much higher than their fuzzy counterparts. Fig. 15 shows the unit step response of the plant $P_2$ along with the load disturbance for ITSE criterion. Both the PID and FOPID have higher overshoot but have a considerably faster settling time. The output of the fuzzy FOPID is very sluggish in this case and the load disturbance suppression is also not very good.

Table3: Optimal parameters for fuzzy FOPID and fuzzy PID controller for plant $P_2$

| Controller type | Performance index | $J^{\min}$ | Controller parameters | | | | | |
|---|---|---|---|---|---|---|---|---|
| | | | $K_e$ | $K_d$ | $\alpha$ | $\beta$ | $\lambda$ | $\mu$ |
| Fuzzy FOPID | ITAE and ISCO | 39.05166 | 0.603307 | 1.142723 | 0.425286 | 2.878081 | 0.996751 | 1.0 |
| | ITSE and ISCO | 35.88284 | 0.012835 | 0.958325 | 0.523293 | 2.059717 | 0.91981 | 0.888057 |
| | ISTES and ISCO | 38.94686 | 0.94137 | 2.147075 | 0.273958 | 1.404629 | 0.982683 | 0.950134 |
| | ISTSE and ISCO | 38.08482 | 0.612212 | 1.175912 | 0.61331 | 3.750646 | 1.118352 | 1.0 |
| Fuzzy PID | ITAE and ISCO | 39.02568 | 0.514478 | 0.945564 | 0.550566 | 3.645142 | - | - |
| | ITSE and ISCO | 37.0897 | 0.156631 | 0.820586 | 0.194024 | 4.604443 | - | - |
| | ISTES and ISCO | 38.63882 | 1.053375 | 2.070359 | 0.218048 | 1.875839 | - | - |
| | ISTSE and ISCO | 38.15205 | 0.988143 | 2.568245 | 0.177589 | 1.765858 | - | - |

Table 4: Optimal parameters for FOPID and PID controller for plant $P_2$

| Controller type | Performance index | $J^{\min}$ | Controller parameters | | | | |
|---|---|---|---|---|---|---|---|
| | | | $K_p$ | $K_i$ | $K_d$ | $\lambda$ | $\mu$ |
| FOPID | ITAE and ISCO | 40.5934 | 2.604385 | 1.610831 | 0.242338 | 0.976384 | 0.604826 |
| | ITSE and ISCO | 46.9178 | 2.812779 | 1.186893 | 0.16216 | 1.195405 | 0.663683 |
| | ISTES and ISCO | 48.64465 | 2.791911 | 2.260489 | 0.30756 | 0.998981 | 0.470879 |
| | ISTSE and ISCO | 47.81009 | 2.606724 | 1.804601 | 0.250482 | 0.999624 | 0.619412 |
| PID | ITAE and ISCO | 46.7316 | 3.401189 | 2.424133 | 0.512058 | - | - |
| | ITSE and ISCO | 45.56443 | 3.04798 | 2.142415 | 0.599768 | - | - |
| | ISTES and ISCO | 46.96438 | 3.717027 | 3.028897 | 0.457318 | - | - |
| | ISTSE and ISCO | 45.95021 | 3.284167 | 2.415988 | 0.554077 | - | - |

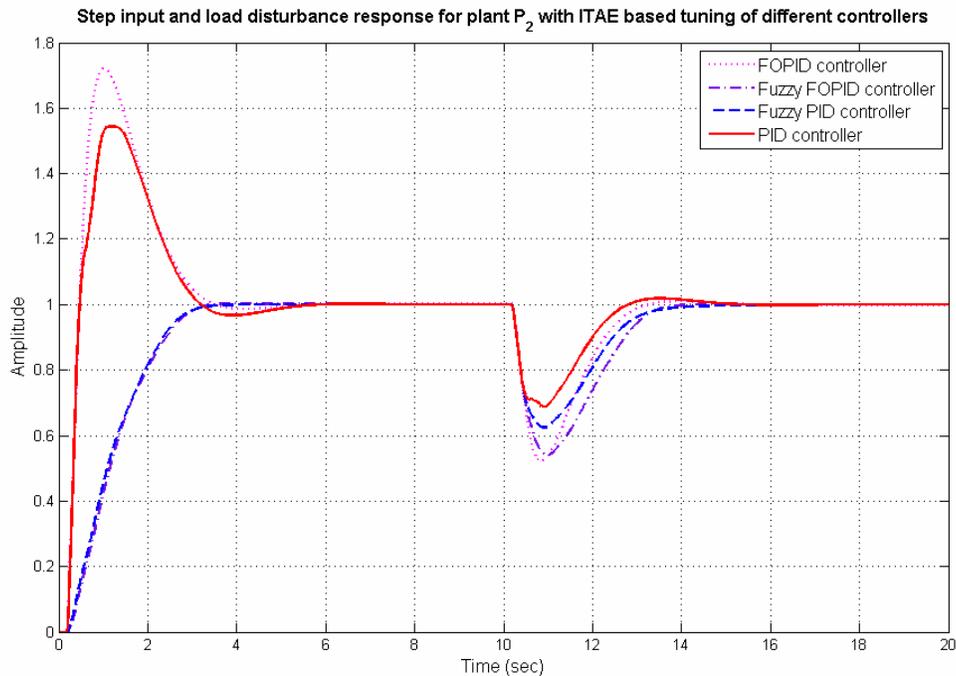

Fig. 13. Step input and load disturbance response for plant $P_2$ with ITAE based tuning.

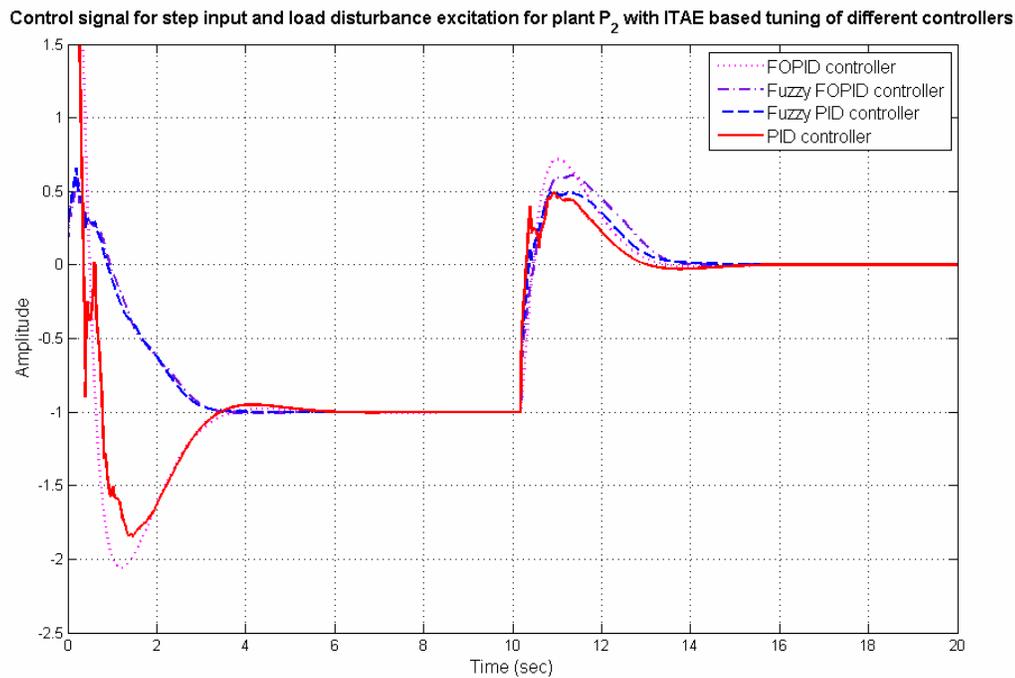

Fig. 14. Control signal of $P_2$ for step input and load disturbance with ITAE based tuning.

Fig. 16 shows the controller output for plant $P_2$ optimized with the ITSE criterion. The initial output of the PID and FOPID controllers is much higher than their fuzzy counterparts.

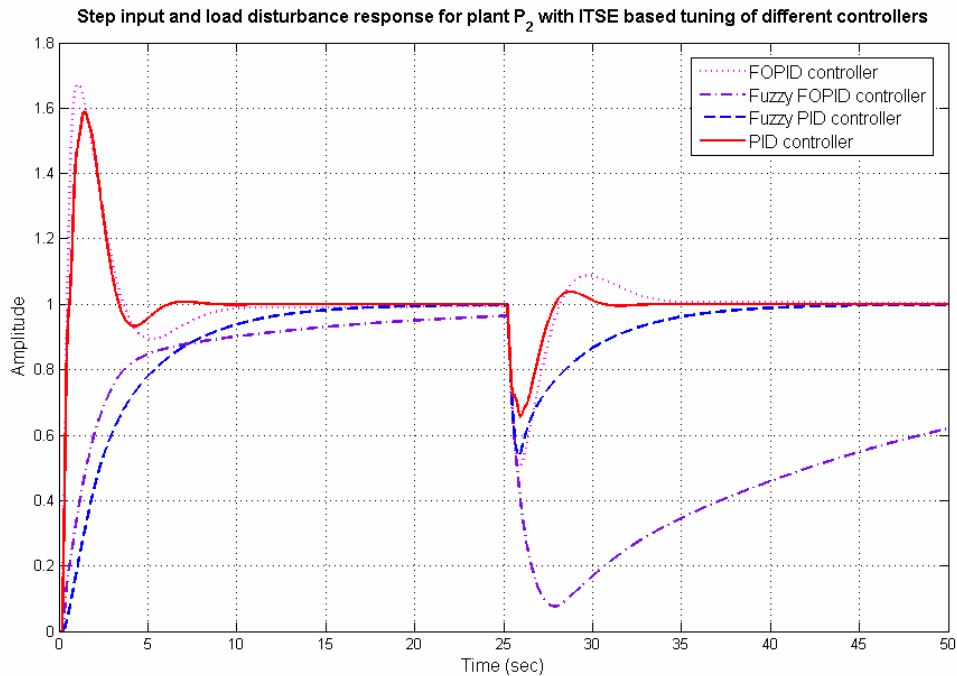

Fig. 15. Step input and load disturbance response for plant $P_2$ with ITSE based tuning.

Fig. 17 shows the unit step response of the same plant along with the load disturbance for the ISTES criterion. The PID and FOPID have a larger peak overshoot. However the load disturbance suppression is better for the PID controller. The fuzzy FOPID controller has a faster rise time compared to the fuzzy PID controller and there is almost no overshoot in both the cases. The load disturbance rejection is almost the same for both the fuzzy PID and the fuzzy FOPID controllers. Fig. 18 shows the controller output for plant $P_2$ optimized with the ISTES criterion. The initial output of the PID and FOPID controllers is much higher than their fuzzy counterparts. Thus the fuzzy controllers would require a smaller actuator size than their PID and FOPID counterparts. Fig. 19 shows the unit step response of plant $P_2$ along with the load disturbance for the ISTSE criterion. The PID and FOPID controller have a higher overshoot than the fuzzy PID and fuzzy FOPID controllers. The load disturbance rejection for the fuzzy PID controller is better than the fuzzy FOPID controller.

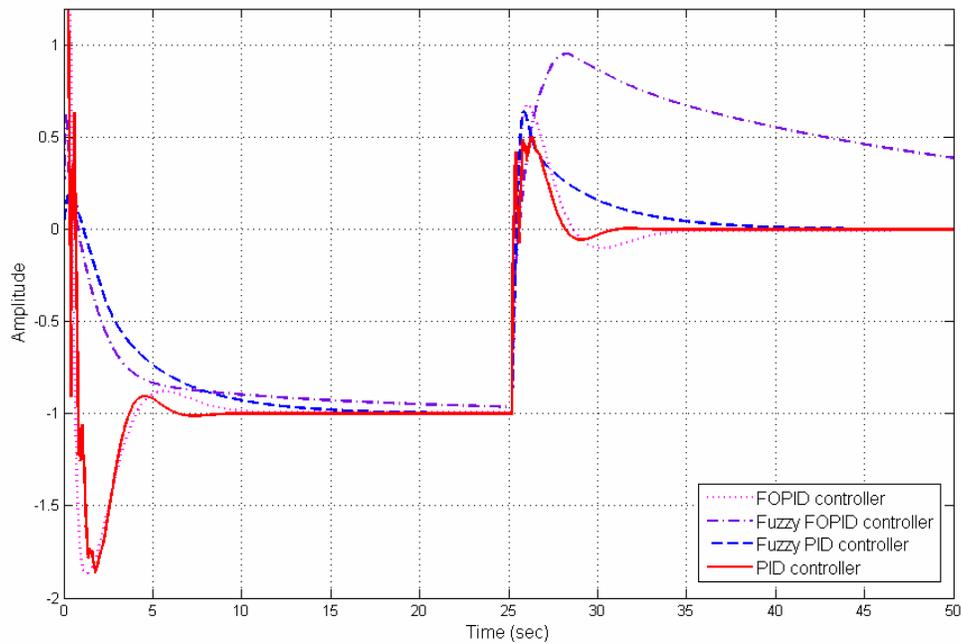

Fig. 16. Control signal of $P_2$ for step input and load disturbance with ITSE based tuning.

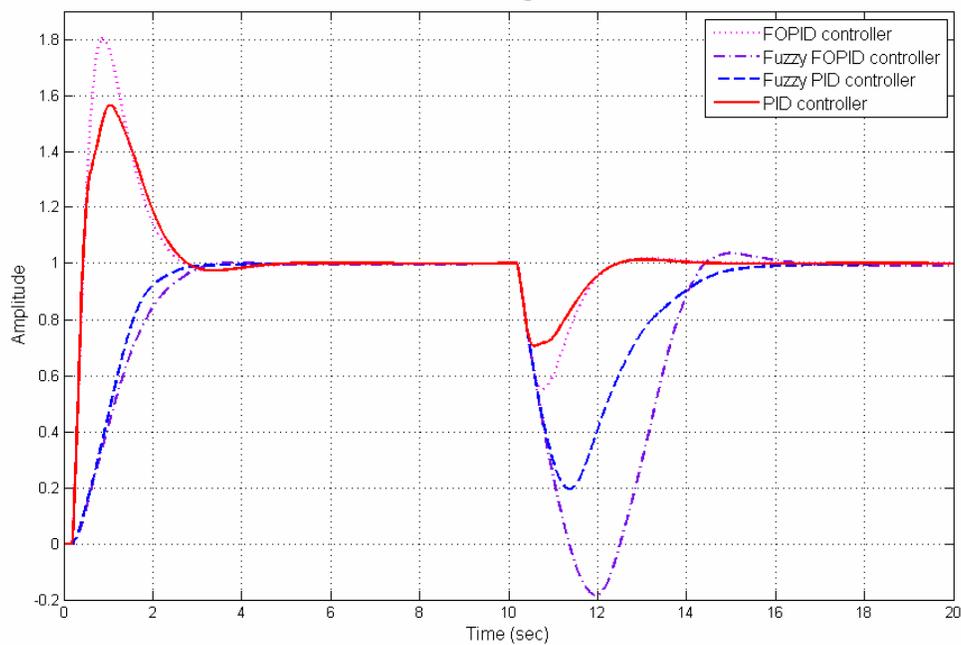

Fig. 17. Step input and load disturbance response for plant $P_2$ with ISTES based tuning.

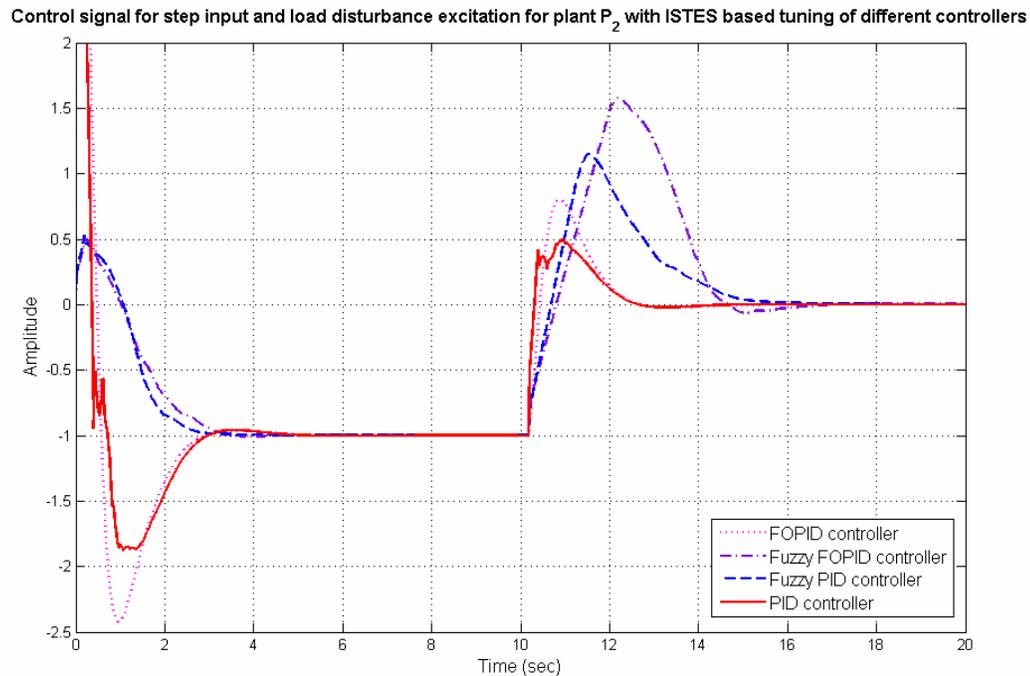

Fig. 18. Control signal of $P_2$ for step input and load disturbance with ISTES based tuning.

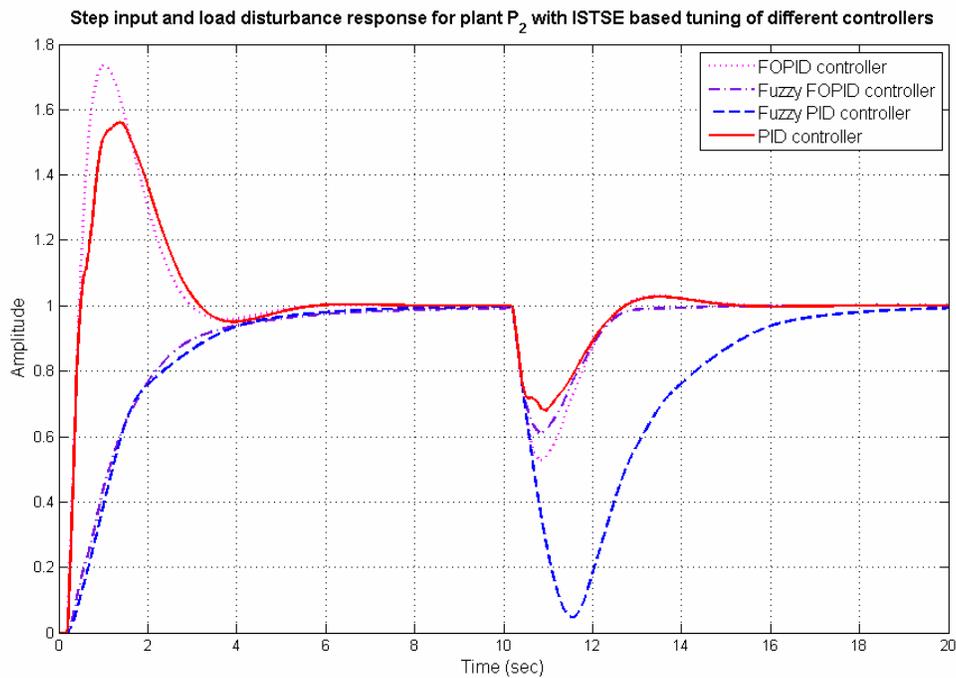

Fig. 19. Step input and load disturbance response for plant $P_2$ with ISTSE based tuning.

Fig. 20 shows the controller output for plant $P_2$ optimized with the ISTSE criterion. The initial control signal of the PID and FOPID controllers is much higher than the fuzzy PID and fuzzy FOPID, which might result in actuator saturation.

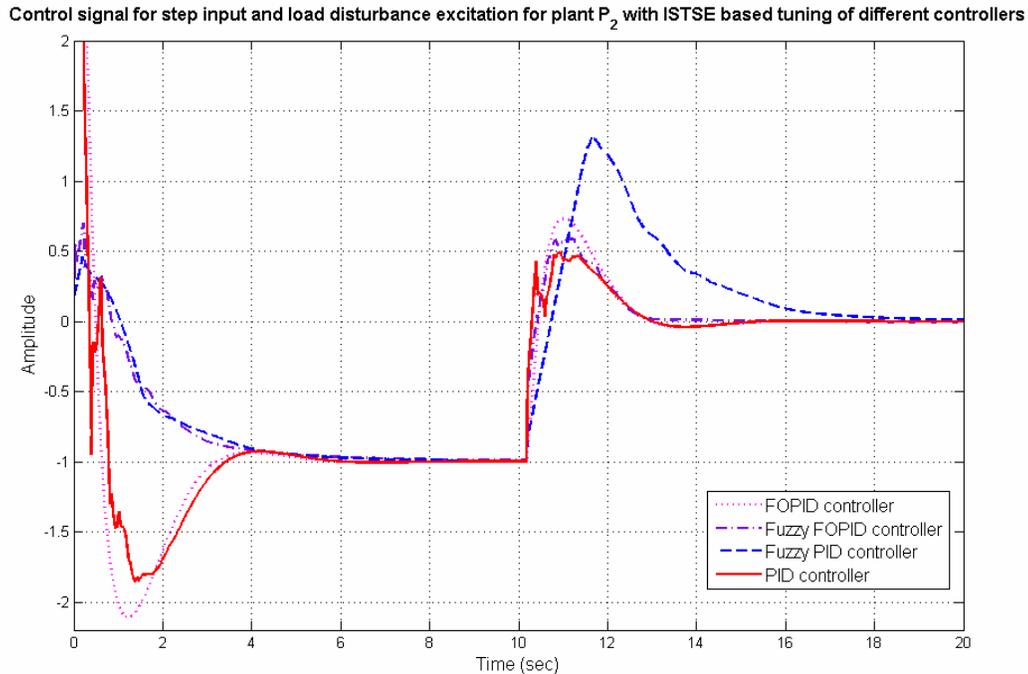

Fig. 20. Control signal of $P_2$ for step input and load disturbance with ISTSE based tuning.

### 4.3. Comparative performance analysis of the different controllers and few discussions

Table 5:
Summary of best controller performances to meet different control objectives for different type of processes

| Type of Process | Performance Index | Best Controller Structure for Different Control Objectives | | |
|---|---|---|---|---|
| | | Set-point tracking | Load-disturbance rejection | Small Control Signal |
| Nonlinear Process with Time Delay | ITAE and ISCO | Fuzzy FOPID | Fuzzy PID | Fuzzy FOPID |
| | ITSE and ISCO | Fuzzy FOPID | Fuzzy FOPID | Fuzzy PID |
| | ISTES and ISCO | Fuzzy FOPID | Fuzzy PID | Fuzzy FOPID |
| | ISTSE and ISCO | Fuzzy FOPID | Fuzzy PID | Fuzzy FOPID |
| Unstable Process with Time Delay | ITAE and ISCO | Fuzzy FOPID | Fuzzy FOPID | Fuzzy PID |
| | ITSE and ISCO | Fuzzy PID | PID | Fuzzy PID |
| | ISTES and ISCO | Fuzzy FOPID | PID | Fuzzy FOPID |
| | ISTSE and ISCO | Fuzzy FOPID | Fuzzy PID | Fuzzy FOPID |

Table 5, lists the best found controller structure from the simulations in a tabular form. It is evident that the proposed Fuzzy FOPID controller outperforms the others for almost all the performance indices for set-point tracking task. However when compared with respect to small magnitude of control signal, the fuzzy PID controller in some cases and fuzzy FOPID in most cases gives better results. Also since the load disturbance attenuation level is not optimized (as the maximum sensitivity specification for linear systems and controllers) by including it in the performance criterion, different controllers

give better results in different cases. Fuzzy controllers are typical nonlinear systems which do not give flexibility to give incorporate sensitivity specification in the design stage unlike linear controllers controlling linear systems and the situation becomes more difficult when the process is inherently nonlinear like the present case. To achieve a comfortable load disturbance level it has been incorporated in the rule base itself describing the loop error and its rate of change as detailed in Mudi & Pal [4].

Also, 30 independent runs (with different seeds for random number generation) were carried out to show the consistency of the GA based controller tuning algorithm. Table 6 reports the mean and standard deviation of the two processes with four controllers each with four different performance indices.

Table 6:
Statistical analysis of the GA based controller tuning results

| Process | Controller | Statistical Measure of the $J_{min}$ for Various Performance Indices | | | | | | | |
| | | ITAE and ISCO | | ITSE and ISCO | | ISTES and ISCO | | ISTSE and ISCO | |
| | | Mean | Standard deviation | Mean | Standard deviation | Mean | Standard deviation | Mean | Standard deviation |
| | PID | 5.842773 | 0.541327 | 4.42718 | 0.001546 | 6.547239 | 0.645439 | 4.667431 | 0.023459 |
| | FOPID | 7.706183 | 0.692537 | 4.538503 | 0.044679 | 13.83508 | 0.518481 | 5.151569 | 0.12559 |
| | Fuzzy PID | 5.420012 | 0.052489 | 4.45654 | 0.005646 | 6.331043 | 0.161723 | 4.705357 | 0.012232 |
| P1 | Fuzzy FOPID | 5.797091 | 0.282726 | 4.470728 | 0.023145 | 7.020855 | 0.470147 | 4.75738 | 0.035402 |
| | PID | 46.74111 | 0.011656 | 45.56801 | 0.003961 | 46.96551 | 0.002417 | 45.95909 | 0.012147 |
| | FOPID | 40.70409 | 0.1894671 | 47.08095 | 0.076367 | 48.9854 | 0.191153 | 47.90995 | 0.128105 |
| | Fuzzy PID | 39.04787 | 0.027437 | 37.1746 | 0.064721 | 38.6766 | 0.043215 | 38.33621 | 0.882041 |
| P2 | Fuzzy FOPID | 39.72096 | 0.712529 | 36.24043 | 0.462656 | 39.36374 | 0.472598 | 38.4043 | 0.711462 |

It is to be noted that the time domain optimality of the fuzzy inferencing process is enforced by the Genetic Algorithm which tunes the various parameters having the higher influence (i.e. the input-output SFs and not the MFs/rule base as discussed earlier) to meet the control performance objectives. Also, it is well known that the PID or FOPID control can be efficiently applied in a control system if the process dynamics is accurately known. Conventional fuzzy logic controller does not rely on the process model since a heuristic control law can be derived from the error and its rate of change. Fuzzy controller gives better performance than conventional PID in the presence of parametric uncertainties, measurement noise and process nonlinearities. Incorporating fuzzy inferencing based PID controller has the both advantages of these two philosophies and has thus been used in the present study to show the control performance enhancement for a delayed nonlinear and an open loop unstable process. Also, the focus of the present paper was on time domain optimal controller tuning since the plant considered for the purpose is a nonlinear one and frequency domain tuning techniques for these types of plants are only available if they are linearized about a certain operating point like [49]. Thus time domain tuning is the preferred method for the tuning of such controllers which works well for a wide variety of processes.

**5. Conclusion:**

   Genetic algorithm based optimal time domain tuning of a novel fractional order fuzzy PID controller is attempted in this paper while minimizing a weighted sum of various integral performance indices and the control signal. Small magnitude of control signal is a necessity in some typical safety critical process control applications like [49] where the chance of actuator saturation and its undesirable results like integral wind-up is highly detrimental and also increases the cost involved for large actuator size as a preventive measure. In the present study four different integral performance indices [43], [47] have been studied while designing the proposed fuzzy FOPID along with its simpler versions like fuzzy PID, $PI^\lambda D^\mu$, fuzzy PID and PID satisfying the same set of optimality criteria. It is observed that the controller performance depends on the type of process to be controlled and also on the choice of integral performance indices. More degrees of freedom in the controller parameters do not necessarily imply better performance in all cases if the performance index is not chosen judiciously. Also for fuzzy enhanced PID controllers it is well known [4] that change in output scaling factor for example has more effect on the controller performance than changes in the membership functions or fuzzification-inferencing-defuzzification mechanism. Thus all the tuning parameters of fuzzy PID controller are not equally potent in affecting the overall performance of the control loop. Our present approach gives additional design parameters viz. the differ-integral orders of a nominal FLC-PID to the designer which can have significant effect on the performance and hence make the applicability of these types of controllers to meet various control objectives.

   Also the performance indices are optimized for set-point change and not for load disturbance in the GA based optimization process. Hence, the optimized controller values are good at set-point tracking, but do not show very good load disturbance rejection response. The comparative study of load disturbance suppression was done for set-point based tuning of optimal controllers [47]-[48]. More stringent multi-objective optimization criteria may be imposed on the controller tuning algorithm to achieve effective results under different circumstances as a scope of future work.

**Acknowledgement:**
This work has been supported by Department of Science and Technology (DST), Govt. of India, under the PURSE programme.